\documentclass[twocolumn,aps,showpacs,superscriptaddress,amssymb]{revtex4}
\usepackage{graphicx}

\begin{document}

\title{Laser cooling of trapped Fermi gases}

\author{Z. Idziaszek}
\affiliation{Institut f\"ur Theoretische Physik, Universit\"at Hannover, D-30167 Hannover,Germany}
\affiliation{Centrum Fizyki Teoretycznej, Polska Akademia Nauk, 02-668 Warsaw, Poland}
\author{L. Santos}
\affiliation{Institut f\"ur Theoretische Physik, Universit\"at Hannover, D-30167 Hannover,Germany}
\author{M. Baranov}
\affiliation{Institut f\"ur Theoretische Physik, Universit\"at Hannover, D-30167 Hannover,Germany}
\affiliation{Russian Research Center Kurchatov Institute, Kurchatov Square, 123182 Moscow, Russia}
\author{M. Lewenstein}
\affiliation{Institut f\"ur Theoretische Physik, Universit\"at Hannover, D-30167 Hannover,Germany}

\begin{abstract}
The collective Raman cooling of trapped one- and two-component Fermi gases is considered. 
We obtain the quantum master equation that describes the laser cooling in the {\it festina lente} regime,
for which the heating due to photon reabsorption can be neglected. For the two-component case the 
collisional processes are described within the formalism of quantum Boltzmann master equation.
The inhibition of the spontaneous emission can be overcome by properly adjusting 
the spontaneous Raman rate during the cooling. Our numerical results based in Monte Carlo simulations 
of the corresponding rate equations, show that three-dimensional temperatures
of the order of $0.08$~$T_F$ (single-component) and $0.03$~$T_F$ (two-component) can be achieved.
We investigate the statistical properties of the equilibrium distribution of the laser-cooled gas, showing 
that the number fluctuations are enhanced compared with the thermal distribution close to the Fermi surface.
Finally, we analyze the heating related to the background losses, concluding that our laser-cooling 
scheme should maintain the temperature of the gas without significant additional losses.
\end{abstract}


\maketitle

\section{Introduction}

The achievement of Bose-Einstein condensation (BEC) \cite{BEC} 
in trapped dilute atomic gases has stimulated a large interest in the 
physics of ultracold gases, both bosonic and fermionic ones.
Related with the latter, similar cooling methods as those employed to reach BEC 
have been recently employed to accomplish a degenerate Fermi gas, i.e. 
a Fermi gas with temperature $T$ below the Fermi temperature ($T_F$) 
\cite{Jin,Schreck,Hulet,Granade,Hadzibabic,Roati}. For $T<T_F$ 
the Fermi pressure becomes noticeable, and as a consequence the Fermi cloud becomes significantly 
broader than a BEC under the same temperature. Additionally, for temperatures below 
a critical one, $T_c$, the system should undergo a Bardeen-Cooper-Schrieffer (BCS) transition 
\cite{DeGennes}.  
When this occurs the system becomes superfluid, due to the Cooper pairing of particles near the 
Fermi surface. 
The accomplishment of BCS pairing and its observation have been considered in detail 
\cite{Stoof, Baranov,Houbiers,Bruun}. However, the employed cooling schemes 
have not yet allowed for temperatures below $T_c$, which has been predicted to be much smaller than $T_F$.
Those cooling methods rely on collisional processes, which in the degenerate regime are strongly suppressed due to 
Pauli blocking. As a consequence, the cooling efficiency decreases, and technical losses have prevented until now from 
reaching temperatures well below $T_F$ \cite{HollandF}. 
Recently, however, it has been proposed that the latter obstacle 
could be overcome by employing Feshbach resonances \cite{Holland,Griffin} or optical lattices \cite{OptLatt} 
in order to increase the ratio $T_c/T_F$ up to $\approx 0.2$, opening the possibility to achieve BCS 
in a temperature range where the collisional processes are still efficient enough. 
In fact, very recently temperatures $T \leq 0.2 T_F$ have been reported by O'Hara {\it et al.} 
\cite{OHara}, which are below the predicted value of $T_C$.

In this paper we show that laser cooling could provide an effective and realistic tool to cool one- and 
two-component Fermi gases well below $T_F$. The proposed method is realized in the  
{\it festina lente} (FL) regime \cite{Festina}, in which the spontaneous emission rate $\gamma$ is smaller 
than the trap frequency $\omega$. In this regime the heating introduced by photon reabsorptions can
be suppressed. In the absence of technical losses the major limitation to the laser cooling 
in the {\it festina lente} regime is set up by the finite value of Lamb-Dicke parameter,
defined as the square root of the ratio of the recoil energy to $\hbar \omega$.
In practice, when the spontaneous emission rate becomes much smaller 
than the trap frequency, the finite available time sets limit on achievable 
temperatures, due to always present loss mechanism.
Reaching FL regime is experimentally feasible, and has been in fact demonstrated in 
experiments by the group of Weiss \cite{Weiss,Salomon}, in which a suppression of 
reabsorptions for decreasing $\gamma/\omega$ has been observed for tightly bound atoms 
in an optical lattice.
The laser cooling in the FL regime down to the quantum degeneracy 
has been already predicted to work for bosons \cite{1atom,Natoms} and 
for polarized fermions \cite{fermicool}. The latter case is particularly interesting, since  
evaporative cooling cannot be employed to a single-component fermionic system, due to the 
absence of $s$-wave scattering, while less restrictive, sympathetic cooling techniques 
\cite{Schreck,Hulet,Hadzibabic,Roati}, are also eventually limited due to Pauli blocking.
The BCS transition in such a system could be realized 
by externally inducing interatomic interactions, for instance dipole-dipole 
ones \cite{dipols}.

In the present paper we analyze the three-dimensional laser cooling of
Fermi gases down to temperatures $T\ll T_F$, by performing Monte Carlo simulations
of the quantum dynamics governed by the Master Equation (ME). 
For $T$ comparable to $T_F$, the laser cooling of fermions is
affected by the statistical inhibition of the spontaneous emission, which
results in a decrease of the cooling efficiency. In a previous work
\cite{fermicool} we have analyzed the laser cooling of a polarized Fermi gas.
We have theoretically predicted that the above mentioned problem
can be overcome by either dynamically adjusting the spontaneous emission
rate in a Raman cooling process, or by employing appropriately designed
anharmonic traps.

We first consider the single-component case. In particular,
we analyze the statistics of
the laser-cooled gases, comparing the results with those expected for a
thermal distribution. We show that whereas the mean-number distribution
can be well approximated by a corresponding thermal one, the
fluctuations significantly differ from the thermal case. This could have
significant effects in the BCS pairing of these gases after an interatomic
interaction is induced in the way discussed above.

In the second part of this work, we analyze the laser-cooling of a
two-component Fermi gas. For this case, the collisions between different
components cannot be neglected. We consider them within the formalism
of Quantum Boltzmann Master Equation (QBME). This formalism is shown
to produced the expected thermal distribution in the absence of laser
cooling. We analyze the case in which only one of the
fermionic species is laser-cooled, whereas the other one is
sympathetically cooled as a result of the collisions between different
components. We show that the laser-cooling is significantly favored by
the presence of the second component. We analyzed also the statistics of
the atom distribution, showing that the fluctuations at the Fermi
surface are significantly enhanced compared to the thermal case.
Finally, we analyze the heating induced by background collisions,
and show that the laser cooling can be employed to maintain a
two-component Fermi gas at a fixed temperature for a relatively long time
without significant atom losses.

The paper is organized as follows. In section II, we discuss our cooling scheme, 
and obtain the ME that describes the Raman cooling 
in the  FL regime. We also describe in this section the corresponding ME for the 
two-component case. We present the resulting rate equations and discuss their 
physical properties. 
In section III we present our numerical results concerning the laser cooling of a single 
component gas, characterizing the statistical properties of the laser-cooled gas.
In section IV we present our results for the case of a two--component Fermi
gas, in which one of the components is laser-cooled. We discuss the statistics for this case, 
as well as the heating induced by losses, and its control using laser-cooling.
We finalize in section V with some conclusions.

\section{Quantum Master Equation} 

We consider $N$ fermionic atoms with an accessible electronic three--level 
$\Lambda$ scheme, with levels $|g\rangle$, $|e\rangle$ and $|r\rangle$. 
We assume that the ground state $|g\rangle$ is coupled via a Raman transition 
to $|e\rangle$ (which is assumed metastable). Another laser couples 
$|e\rangle$ to the upper state $|r\rangle$, which rapidly decays into $|g\rangle$. 
After the adiabatic elimination of $|r\rangle$, a two--level system is 
obtained, with an effective Rabi frequency $\Omega$, and an effective 
spontaneous emission rate $\gamma$. The latter can be controlled 
by modifying the coupling from $|e\rangle$ to $|r\rangle$. 
The atoms are confined in a non-isotropic dipole trap with  
frequencies $\omega^g_{x,y,z}$, $\omega^e_{x,y,z}$, different for the 
ground and the excited states, and non-commensurable one with another. 
The latter assumption simplifies enormously the dynamics of the 
spontaneous emission processes in the FL limit. When deriving the quantum ME, 
it is important to cancel the off-diagonal terms of the density matrix, in order 
to reduce the ME to a rate equation, which is possible to 
simulate numerically with Monte Carlo methods. The anisotropy of the trap 
does not need to be very large, and for practical purposes can be rather small. 
Its effects in the rate equations is not relevant at all, and consequently in our simulation we consider 
the simple case of isotropic trap. The cooling process consists of 
sequences of Raman pulses of appropriate frequencies, adjusted is such a way that they 
induce the transition of atoms to the lower motional states of the trap.
We assume weak optical excitation, so that
no significant population in $|e\rangle$ is present. This allows to
adiabatically eliminate $|e\rangle$, and
consequently to consider only the density matrix
$\rho(t)$ describing all atoms in $|g\rangle$, and 
being diagonal in the Fock representation corresponding to the bare 
trap levels.

Let us introduce the annihilation and creation operators of atoms in the ground (excited) state
and in the trap level $m$ ($l$), which we denote by $g_{m}$, $g_{m}^{\dag}$ ($e_{l}$, $e_{l}^{\dag}$).
These operators fulfill standard fermionic anticommutation relations: 
$\{g_{m},g_{n}^{\dag}\}=\delta_{m,n}$, $\{e_{m},e_{n}^{\dag}\}=\delta_{m,n}$. We use the standard 
theory of quantum-stochastic processes \cite{GardinerB,Carmichael} to develop  
the quantum ME that governs the dynamics of cooling \cite{Castin}.
The derivation of the ME for fermions 
proceeds basically in a similar way to the corresponding derivation for bosons. The latter was 
presented in details in \cite{Natoms}. The ME that describes the dynamics of the density 
matrix $\rho$ of the atoms in the FL regime is
\begin{equation}
\dot\rho(t)={\cal L}_0\rho+{\cal L}_1\rho,
\label{ME}
\end{equation}
where
\begin{eqnarray}
{\cal L}_0\rho&=&-i\hat H_{eff}\rho(t)+i\rho(t)\hat H_{eff}^{\dag}+{\cal J}\rho(t), \\
{\cal L}_1\rho&=&-i[\hat H_{las},\rho(t)], 
\end{eqnarray}
with
\begin{eqnarray}
\hat H_{eff}&=&\sum_{m} \omega_{m}^{g}g_{m}^{\dag}g_{m}+\sum_{l}(\omega_{l}^{e}
-\delta)e_{l}^{\dag}e_{l} \nonumber \\
&-& i \gamma \sum_{l,m} \xi_{lm} e_{l}^{\dag}g_{m}g_{m}^{\dag}e_{l}, \label{Heff}\\
\hat H_{las}&=&\frac{\Omega}{2}\sum_{l,m}\eta_{lm}(k_{L})e_{l}^{\dag}g_{m}+H.c.,\\
{\cal J}\rho(t)&=& 2 \gamma \sum_{l,m}\xi_{lm} g_{m}^{\dag}e_{l}\rho(t) e_{l}^{\dag}g_{m}.   
\end{eqnarray}
Here, $2\gamma$ is the single--atom effective spontaneous emission rate,
$\Omega$ is the  Rabi frequency associated with the atom transition and the laser field, 
$\omega_m^g$  ($\omega_l^e$) are the energies of the ground (excited) harmonic trap
level $m$ ($l$), $\delta$ is the laser detuning from the atomic transition,
$\xi_{lm}=\int_{0}^{2\pi}d\phi\int_{0}^{\pi}d\theta \, \sin\theta \, {\cal W}(\theta,\phi) |\eta_{lm}(\vec{k})|^2$,
where ${\cal W}(\theta,\phi)$ is the fluorescence dipole pattern, and   
$\eta_{lm}(k_{L})=\langle e,l|e^{i\vec k_{L}\cdot\vec r}|g,m\rangle$
are the Frank--Condon factors.
In the following we assume that $\Omega<\gamma$ and $\Omega^2/\gamma \ll \omega$, which allows for 
adiabatic elimination of excited state $|e\rangle$. To assure the FL regime we require also 
$\gamma < \omega$ at least in the initial phase of the cooling process.

The dynamics of the two-component system in the FL limit, where only the first component is laser cooled, 
may be described by the following ME: 
\begin{equation}
\dot\rho(t)={\cal L}_0\rho+{\cal L}_1\rho+{\cal L}_2\rho.
\label{ME1}
\end{equation}
The new term ${\cal L}_2$ that appears in Equation (\ref{ME1}) describes the process of elastic collisions 
between the two species:
\begin{eqnarray}
{\cal L}_2\rho&=&-i[\hat H_{coll},\rho(t)], 
\end{eqnarray}
where
\begin{equation}
\hat H_{coll}=\sum_{n,m,q,p} U_{n,m;q,p} g_{n}^{\dag} b_{m}^{\dag} b_{q} g_{p}.
\label{Hcoll}
\end{equation}
Here, $b_{m}$ and $b_{m}^{\dag}$ are respectively the annihilation and creation operators of atoms 
of the second component in the trap state $m$. 
Since at low temperatures collisions between atoms are purely of $s$-wave type,
the operator $\hat H_{coll}$ accounts only for collisions
between atoms in the second component and those atoms in the ground electronic state $|g\rangle$ 
of the first species. The collisions with the atoms in the $|e\rangle$ can be neglected
if the laser is sufficiently weak to assure that only a small fraction of atoms is 
excited in each pulse. The amplitude $U_{n,m;p,q}$ in Equation 
(\ref{Hcoll})
reads
\begin{equation}
U_{n,m;q,p}=\frac{4\pi\hbar a_{sc}}{m}\int_{R^3}d^3x
\phi_{n}^{\ast}({\bf x}) \beta_{m}^{\ast}({\bf x}) \beta_{q}({\bf x}) \phi_{p}({\bf x}),
\label{Udef}
\end{equation}
where $\phi_{n}$ ($\beta_{n}$) denotes the wavefunction of the $n$-th level of the $|g\rangle$, 
($|b\rangle$) trap, and $a_{sc}$ is the scattering length.
The term ${\cal L}_1$ from the ME (\ref{ME1}), includes the effective Hamiltonian $H_{eff}$,
which in the presence of two components, accounts also for the energies of atoms of the second 
component
\begin{eqnarray}
\hat H_{eff}&=&\sum_{m}\omega_{m}^{g}g_{m}^{\dag}g_{m}+\sum_{l}(\omega_{l}^{e}
-\delta)e_{l}^{\dag}e_{l} + \nonumber \\
& + & \sum_{n}\omega_{n}^{b} b_{n}^{\dag}b_{n} 
- i \gamma \sum_{l,m} \xi_{lm} e_{l}^{\dag}g_{m}g_{m}^{\dag}e_{l}. \label{Heff2}
\end{eqnarray}
For the assumed regime of parameters, the excited state may be adiabatically eliminated.
To this end we use standard projection operator techniques \cite{GardinerB}. The corresponding 
calculus proceed in the same way as in the case of bosons,  
which was described in detail in \cite{Natoms}. Here, we present only the final results. 
For sufficiently weak interactions between the two species, the dynamics due to the collisions and 
the dynamics due to the laser--cooling are independent \cite{Coll}. The process of laser cooling 
may be described by the ME (\ref{ME}) without the collisional part, while the collisional part of dynamics
after the adiabatic elimination takes the form of QBME, similar 
to the one formulated for bosons and presented in \cite{Gardiner1, Gardiner2}. 
The rate equations for the populations $N_n^{1}$ and $N_m^{2}$ in level $n$ and $m$ of the trap $|g\rangle$ 
and $|b\rangle$ respectively, are given by
\begin{eqnarray}
\dot{N}_n^{1} & = &\sum_{p} P^{opt}_{p\rightarrow n} N_p^{1} + \sum_{m,p,q} P^{coll}_{p,q \rightarrow n,m} N_p^{1} N_q^{2}
\nonumber \\
& & \mbox{} - \sum_{p}  P^{opt}_{n\rightarrow p} N_n^{1} - \sum_{m,p,q} P^{coll}_{n,m \rightarrow p,q} N_n^{1} N_m^{2}
, \label{RateEq1} \\
\dot{N}_m^{2} & = & \sum_{n,p,q} P^{coll}_{p,q \rightarrow n,m} N_p^{1} N_q^{2} \nonumber \\ 
& & \mbox{} - \sum_{n,p,q} P^{coll}_{n,m \rightarrow p,q} N_n^{1} N_m^{2},
\label{RateEq2}
\end{eqnarray}
where $P^{opt}_{p\rightarrow n}$ is the rate of laser-induced transition from state $m$ to state
$n$ of  the $|g\rangle$ trap 
\begin{eqnarray}
P^{opt}_{p\rightarrow n} &=&  \frac{\Omega^{2}}{2\gamma} (1-N_n^{1}) \times \nonumber \\
\times & \sum_{l} & \frac{\gamma^2 \xi_{ln} \eta_{lp}(k_{L})}
{[\delta-(\omega_l^e-\omega_p^g)]^2+\gamma^2 (R_{l}+\xi_{lp})^2},
\label{Pmn}
\end{eqnarray}
with $R_{l} = \sum_{n'} \xi_{ln'} (1-N_{n'}^{1})$. The rate $P^{coll}_{p,q \rightarrow n,m}$
of a collision between two fermions of different species, occupying 
the states $p$ and $q$ of the trap respectively, resulting in a transition to the states 
$n$ and $m$, is given by   
\begin{equation}
P^{coll}_{p,q \rightarrow n,m} = \frac{\pi}{\omega} 
\left|U_{n,m;q,p}\right|^2 \delta_{E_n+E_m,E_p+E_q} (1-N_n^{1}) (1-N_m^{2}).
\label{Pcoll}
\end{equation} 
In the transition rate (\ref{Pmn}) one can clearly identify two contributions due to
the quantum--statistical character of fermions: (a)   
If the system becomes more degenerate, $R_{l}$ vanishes, and the atoms remain for very long 
times in the excited state $l$, i.e. the spontaneous emission is inhibited. This 
affects negatively the cooling process in two ways: first, it prolongs the cooling, and 
second, the excited-ground collisions can occur and lead to heating and losses.
In addition, the adiabatic elimination used in the derivation of 
Equation. (\ref{Pmn}) ceases to be valid. (b) The fermionic inhibition factor $(1-N_{n})$ appears in 
the numerator of the probabilities, introducing also a slowing-down of the cooling process.

The negative influence of Fermi statistics above can be overcome by dynamically changing of $\gamma$ 
in Raman cooling \cite{fermicool}. One can increase it gradually during the cooling in order to avoid 
the inhibition effects, but still remaining in the FL regime. Note that, the FL condition involves 
the spontaneous emission rate modified in the presence of other atoms: $\gamma R_{l}< \omega$. 
Still, even if one uses such approach some small fraction of the atoms will remain 
 in the excited state after the cooling pulse, and has to be removed from the trap in order to avoid 
non-elastic collisions. The latter aim can be achieved by optically pumping the 
excited atoms to a third non-trapped level. This introduces a new loss mechanism 
which we take into account in the simulations.

The calculations of the laser cooling were performed for the case of a three-dimensional harmonic trap. 
For simplicity
we assumed that the trap is isotropic with frequency $\omega$. The main difficulty 
for performing Monte--Carlo simulations of the laser--cooling for fermions is the large number of states
in the calculation. Since one state can be populated maximally by one fermion, 
in order to perform a simulation for a medium-size atomic sample (in our case $10^4$ atoms), we had to 
consider rather large trap ( $80$ energy levels in our case). In addition if the simulation starts 
from a temperature above $T_F$, the number of states must be much larger than the number of atoms in 
the trap. For an initial number of $10^4$ atoms, we calculated the cooling dynamics for the trap consisting
of approximately $10^5$ states. 
For a single component gas of fermions, where the collisions between the atoms are
absent, we were able to perform simulations treating each state of the trap separately.
In this case the transition is only induced by the laser, and its rate given by (\ref{Pmn}) depends
only on two states. The number of different probabilities that one have to recalculate after the change
of the atomic distribution is roughly $10^9$. 

In the case of the two--component Fermi gas, one has to account for the process of collisions, 
in which the transition rate (\ref{Pcoll}) depends on four
numbers: two initial states and two final states. The number of different probabilities that one has to
recalculate after a single transition takes place is roughly $10^{18}$. The number of operations required to 
calculate
the probabilities is orders of magnitude too large for numerical computations. Therefore for simulating
the cooling in two-species gas, we have applied the ergodic approximation. We assume that the populations
of the states with the same energy are equal: $N_m^{n}=\bar{N}_{E(m)}^{n}/g_{E(m)}$, where 
$n\in\{1,2\}$, $\bar{N}_{E}^{n}$ is the number of atoms of the $n$-th component
occupying the energy shell $E$ and $g_{E}$ is the
degeneracy of the energy shell $E$. This approximation relies on the fact 
that for typical parameters, the collisional processes are much faster than the laser cooling.
Additionally the thermalization inside a single energy shell is much faster than the thermalization 
between energy shells of different energies.

The assumption that collisions occur much more frequently than laser induced transitions, is necessary,
since the laser cooling is not fully ``compatible'' with the ergodic approximation. The cooling is based
on the process of absorption of a photon from one of three orthogonal laser beams, therefore the rate 
of transition depends on the three dimensional structure of the initial and the final state. 
Due to this 
dependence, we have not been able to find a simpler formula, which would be a counterpart of 
Equation (\ref{Pmn}) in ergodic approximation. In the case of the collisional process, however, 
one can further develop the rate (\ref{Pcoll}) in order to fully utilize the advantages of the 
ergodic approximation. 
 
The rate of collision, between fermions occupying the energy shells $e_1$ and $e_2$, 
resulting in a transition to the energy shells $e_3$ and $e_4$ in the ergodic approximation is given by
\begin{equation}
P^{coll}_{e_1, e_2 \rightarrow e_3,e_4} = 
\frac{\pi}{\omega} \frac{ (g_{e_3}-\bar{N}_{e_3}^{1}) 
(g_{e_4}-\bar{N}_{e_4}^{2})}
{g_{e_1} g_{e_2} g_{e_3} g_{e_4}} \left|\tilde{U}_{e_1,e_2;e_3,e_4}\right|^2,
\label{Pcollerg}
\end{equation}
where the amplitude of the transition $\tilde{U}_{e_1,e_2;e_3,e_4}$ is defined as follows
\begin{eqnarray} 
\left|\tilde{U}_{e_1,e_2;e_3,e_4}\right|^2 & = & \sum_{n,m,p,q} 
\delta_{E_p,e_1} \delta_{E_q,e_2} \delta_{E_n,e_3} \delta_{E_m,e_4} \times \nonumber \\
& \times & \left|U_{n,m;q,p}\right|^2.
\label{Uerg}
\end{eqnarray}
In principle, the amplitude of transition  $\tilde{U}_{e_1,e_2;e_3,e_4}$ may be calculated directly from
Equation. (\ref{Uerg}), by performing a summation of the terms $\left|U_{n,m;q,p}\right|^2$ given by   
formula (\ref{Udef}). This requires, however, a lot of numerical efforts, since the number of operations
is of the order $N_{st}^4$, where $N_{st}$ is the number of states in the trap. This can be avoided, 
if the trap is isotropic, in which case we were able to derive a simplified expression
\begin{eqnarray}
\left|\tilde{U}_{e_1,e_2;e_3,e_4}\right|^2 &=& \frac{2}{\pi^4} 
\left( \frac{4 \pi \hbar a_{sc}}{m \xi^3} \right)^2 \times \nonumber \\
& \times & \sum_{n,m,p,q} \frac{I_{nmpq}^{4} I_{nmpq}^{0} - (I_{nmpq}^2)^2}{2^{n+m+p+q} n! m! p! q!},
\label{Uerg1}
\end{eqnarray}
where $\xi=\sqrt{\hbar/m\omega}$. The coefficients $I_{nmpq}^k$ are defined by 
\begin{equation}
I_{nmpq}^k = \int_{- \infty}^{\infty} dx e^{-x^2} x^k H_n(x) H_m(x) H_p(x) H_q(x), 
\end{equation} 
where $H_n(x)$ denotes the $n$-th Hermite polynomial. Note that the summation over indices $n$ ($m$, $p$, $q$)
in Equation (\ref{Uerg1}), runs from $0$ if $e_n$ ($e_m$, $e_p$, $e_q$)  is even, or  
from $1$ if $e_n$ ($e_m$, $e_p$, $e_q$)  is odd, up to $e_n$ ($e_m$, $e_p$, $e_q$) 
with an increment equal to $2$. In order to obtain 
Equation (\ref{Uerg1}), we have used standard summation formulas for Hermite polynomials \cite{Ryzhik}. 
Equation (\ref{Uerg1}) can
be further simplified, if the collisions involves distant energy shells, i.e. the minimum $e_{min}$
of the set of numbers $\{e_1,e_2;e_3,e_4\}$ is much smaller than its maximum $e_{max}$. According to Ref. 
\cite{HollandB}, the following approximate formula holds
\begin{equation}
\left|\tilde{U}_{e_1,e_2;e_3,e_4}\right|^2 \simeq \left( \frac{4 \pi \hbar a_{sc}}{m \xi^3} \right)^2
\frac{g_{e_{min}}}{4 \pi^4}.
\label{Uappr}
\end{equation}
With the help of the exact expression (\ref{Uerg1}), we have verified numerically that the approximation 
(\ref{Uappr}) gives correct results, already for $e_{max}-e_{min} \gtrsim 2 \hbar \omega$. 
Equation (\ref{Uappr}) leads to small errors for low lying energy shells, and for the transitions between 
neighboring shells. In the simulations, however, we have used the values calculated numerically with
the help of the exact formula (\ref{Uerg1}).

\section{Results for one--component gas}

The numerical simulations of the cooling were performed using Monte Carlo techniques. In the calculation we
assumed that the atoms are confined in a dipole trap, characterized by a Lamb-Dicke parameter 
$\eta=2\pi a/\lambda=2$, with $a=\sqrt{\hbar/2m\omega}$ being the size of the ground state of the trap, and 
$\lambda$ the laser wavelength. In the context of laser cooling, an optical dipole trap was recently  
employed in experiments on all-optical
production of a degenerate gas of two species of Li \cite{Granade}, and it is also considered in current 
experiments in Mg \cite{Ernst}. Due to numerical limitations we assumed Lamb-Dicke parameter 
$\eta = 2$ \cite{LambDicke}. The assumed value of $\eta$ is not unrealistic, and can be realized for instance by trapping 
potassium atoms in a dipole trap with $\omega = 2 \pi \times 2.4$~kHz, which is the trapping frequency 
employed in experiments in $Li$ \cite{Granade}, and a laser wavelength $\lambda \simeq 720$~nm.
In the calculation we use the above mentioned values to calculate real cooling 
time. We start our simulations with an initial number of atoms $N=10660$, which corresponds
to a Fermi energy $E_F = 38 \hbar \omega$. The three-dimensional isotropic harmonic trap contains $81$ 
energy levels, which corresponds to $91881$ states.

In the gas of fermions the main sources of losses are background collisions and photoassociation.  
Background collisions result from non ideal vacuum conditions in the experiment, and 
introduce a relevant heating mechanism for temperatures well below Fermi temperature \cite{Timmer}.
In the simulations we assumed a background collision rate $\gamma_{bg} = 1/350$ Hz \cite{HollandF}.
Photoassociation losses (when the laser is tuned between molecular resonances) 
are typically of the order of $10^{-14}$cm$^3$/s for 
laser intensities of $1$mW/cm$^2$ \cite{Machholm}. In our case the laser intensities needed 
are typically $1000$ times smaller and we estimate that for $N=10660$ atoms, the atomic density is
smaller than $3.5 \times 10^{14}$cm$^{-3}$. Therefore the photoassociation losses can be safely neglected.
The other type of losses which we account for in our calculations is the removal process of long-living 
excited atoms at the end of each cooling pulse.

\begin{figure}[tb]
\includegraphics[width=8cm,clip]{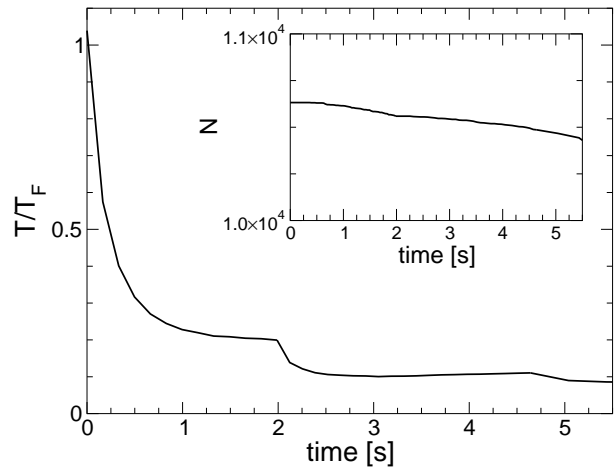}
\caption{
\label{fig:1}
Time-dependence of the temperature during the laser 
cooling of a single component Fermi gas by controlling the effective spontaneous Raman rate. 
Inset: time dependence of the number of atoms, which decreases due to background collisions and 
the removal of long-living excited atoms. 
The different Raman cooling stages discussed in the text extend from $0$~$s$ to $2$~$s$, 
from $2$~$s$ to $4.6$~$s$, and from $4.6$~$s$ to $5.5$~$s$.}
\end{figure}

Figure~\ref{fig:1} shows the time-dependence of the temperature of a laser-cooled one-component gas 
of fermions. The inset shows the time dependence of the number of fermions, that decreases due to losses.
As the initial state
of the system we consider a thermal distribution with temperature $T_0=1.04 T_F$.
The cooling process was divided into three stages, each consisting of a sequence of two Raman pulses.
  The employed pulses that are used are characterized by the following parameters: detuning   
$\delta/\omega=\{(-11,-12)$, $(-16,-17)$, $(-19,-20)\}$ respectively, Rabi frequency
$\Omega/\gamma=\{(0.013,0.013)$, $(0.007,0.01)$, $(0.1,0.8)\}$ respectively, and length
$\Delta t/\omega^{-1}=\{(250,250)$, $(500,500)$, $(2000,4000)\}$ respectively. 
The use of two pulses with slightly different detunings in each cooling stage allows to 
compensate for the oscillatory character of the Frank-Condon factors.
The values of $\Omega$
were chosen in such way that not more than $10\%$ of the atoms is excited during each pulse, in order 
to fulfill the condition of the adiabatic elimination. The temperature of the system was determined by 
fitting of the thermal distribution to the current distribution of fermions. At the end of cooling,
the system reaches the temperature $T=0.08 T_F$.

\begin{figure}[tb]
\includegraphics[width=8cm,clip]{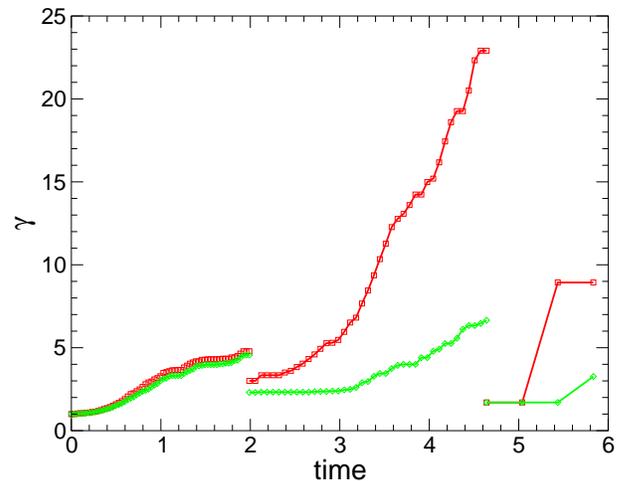}
\caption{
\label{fig:2}
Time-dependence of the 
effective single-atom spontaneous emission rate for the case considered in Fig.~\ref{fig:1}.
The values for the first pulse in the sequence (solid line with diamonds)
and those for the second pulse (solid line with squares) are indicated.
The different Raman cooling stages discussed in the text extend from $0$~$s$ to $2$~$s$, 
from $2$~$s$ to $4.6$~$s$, and from $4.6$~$s$ to $5.5$~$s$.}
\end{figure}

\begin{figure}[b]
\includegraphics[width=8cm,clip]{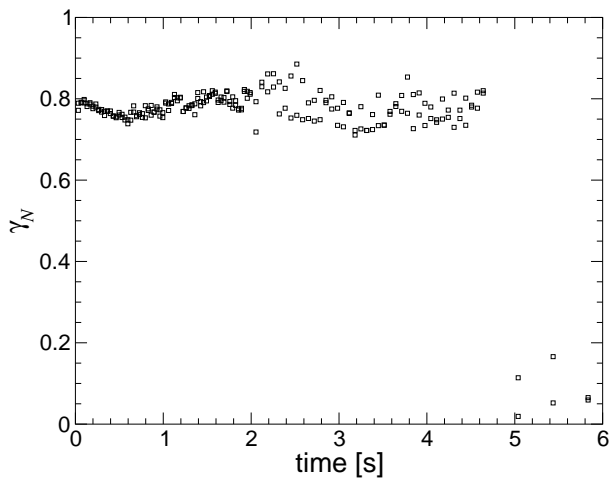}
\caption{
\label{fig:3}
Time-dependence of the expectation value of the spontaneous emission rate for the 
case considered in Fig.~\ref{fig:1}.
The spontaneous emission rate $\gamma_N$ takes into account the inhibition due to the presence of the
other fermions in the trap. The average was calculated for the atomic distribution after each 
cooling pulse. The different Raman cooling stages discussed in the text extend from $0$~$s$ to $2$~$s$, 
from $2$~$s$ to $4.6$~$s$, and from $4.6$~$s$ to $5.5$~$s$.}
\end{figure}

The time dependence of the spontaneous emission rate is presented in Figure~\ref{fig:2}. The two curves 
depict the values for the two pulses used at each stage of cooling. 
From this figure one can see how $\gamma$ was controlled in order to speed up the cooling, and at the 
same time avoid inhibition of the spontaneous emission, and remain in the {\it festina lente} regime. In
the simulation it was realized by increasing of $\gamma$ gradually during the cooling process, 
in such a way that the average value of 
the spontaneous emission rate calculated for many-atom system: $\gamma_N$, remains approximately constant, 
and equal to $0.8\omega$, except for the last pulses. In the last cooling stage  rather long pulses 
were applied, and therefore $\gamma_N$ could be considered smaller than in the other pulses. 
The expectation value of 
$\gamma_N$ is presented on Figure~\ref{fig:3}. It was calculated at the end of each pulse, averaging over
all possible laser-induced transition that may take place. With each transition $m \rightarrow n$ 
one can associate the spontaneous emission rate $\gamma_N(l) = \gamma R_{l}$, where $l$ denote an 
intermediate state, when the atom is excited.

\begin{figure}[ht]
\includegraphics[width=8cm,clip]{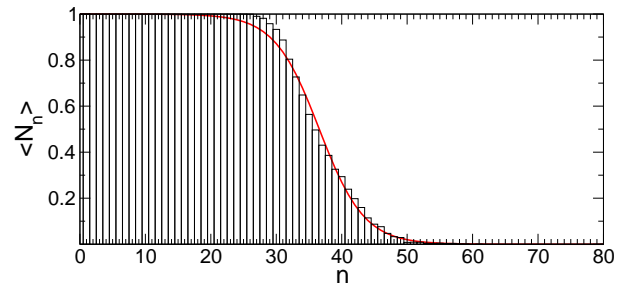}
\caption{
\label{fig:4} 
Mean distribution of fermions (bars) obtained by averaging in time during the prolonged 
last stage of the cooling process
presented in Fig.~\ref{fig:1}. Solid line - fit of the thermal distribution.  
}
\end{figure}

\begin{figure}[ht]
\includegraphics[width=8cm,clip]{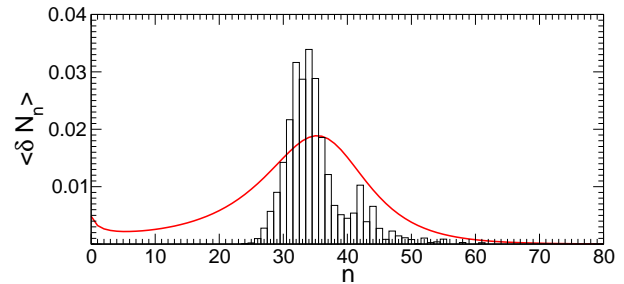}
\caption{
\label{fig:5}
Fluctuations of the distribution of fermions (bars) 
obtained by averaging in time during the prolonged last stage of the cooling 
process presented in Fig.~\ref{fig:1}. Solid line - fluctuations of the thermal gas, 
with the temperature and the chemical 
potential determined from the fit of the mean distribution. 
}
\end{figure}

At the end of the cooling process, the temperature of the gas of fermions stabilizes and the 
system reaches an equilibrium distribution. In the absence of collisions, this distribution is purely
determined by the laser--induced transitions, and in general may be very different than the thermal 
distribution. In order to investigate this point, we studied the mean population and the fluctuations 
of the energy shells. The statistics was calculated by time-averaging during the last stage of cooling.
To this end, the final stage of cooling was extended to the time $t=4 \times 10^{5} \omega^{-1}\approx 26 s$.
This time correspond, in average to four absorption and spontaneous emission cycles for each atom.
The average distribution of atoms is plotted in Figure~\ref{fig:4}. The bars depict the distribution
obtained from the simulation, whereas the solid line shows the fit of the thermal distribution.
According to Figure~\ref{fig:4}, the mean population numbers are well described in the framework
of the grand canonical ensemble. 
The fluctuations of the occupation numbers are presented in Figure~\ref{fig:5}. The solid line as before
represents the fluctuations of the thermal gas, with the temperature and the chemical potential determined
from the fit of the mean distribution. As can be seen, the fluctuations in the laser cooled gas have
completely different properties than the fluctuations in the thermal distribution. 
In a small energy range below 
$E_F$, the fluctuations are enhanced, whereas deeply in the Fermi sea they completely vanish. 
From this peculiar behavior we may conclude that the laser cooling strongly affects the atoms 
near the Fermi surface. On the other hand, it does not excite those atoms deeply in the Fermi sea, due 
to the large negative detunings of the employed pulses.

The existence of such non-thermal fluctuations is a peculiar feature of the proposed cooling method.
The observed enhancement of the fluctuation near the Fermi energy could have an important influence 
on the formation of Cooper pairs in laser cooled Fermi gases. 
This question, however, lies beyond the scope of this paper and 
it will be the subject of further investigations. It is worth stressing that our cooling scheme 
can be applied to a Fermi mixture with scattering length
modified in the vicinity of a Feshbach resonance. With the help of this technique, 
the critical temperature for the BCS transition is predicted to reach $0.2$-$0.5$~$T_F$ 
\cite{Holland, Griffin}. Laser cooling
should then allow to achieve such temperatures, presumably in a time scale shorter that $1$~$s$
(see Figure~\ref{fig:1}).

\section{Results for two--component gas}
 
In the simulation of laser cooling in the two--component Fermi system we considered the same parameters 
as in the case of the single component gas. We assumed that the two species are placed in the
same isotropic optical trap with frequency $\omega = 2 \pi \times 2.4$~kHz. 
The Lamb-Dicke parameter is $\eta=2$ and the wavelength of the cooling laser is $\lambda \simeq 720$~nm.
The trap considered in the simulation has the same depth for both species, and contains $81$ energy levels.
The losses due to the background collisions affect in this case both components. In the simulations 
we assumed the same background-collision rate $\gamma_{bg} = 1/350$ Hz for both species. The two components 
have equal initial number of atoms $N=10660$. The scattering length for the interactions
between the two species is equal to $a_{sc}=157 a_0$, where $a_0$ is the Bohr radius. This
value correspond to the interactions between two spin states $|F=9/2,m_F=9/2\rangle$ and
$|F=9/2,m_F=7/2\rangle$ of $^{40}$K \cite{DeMarco}.

\begin{figure}[b]
\includegraphics[width=8cm,clip]{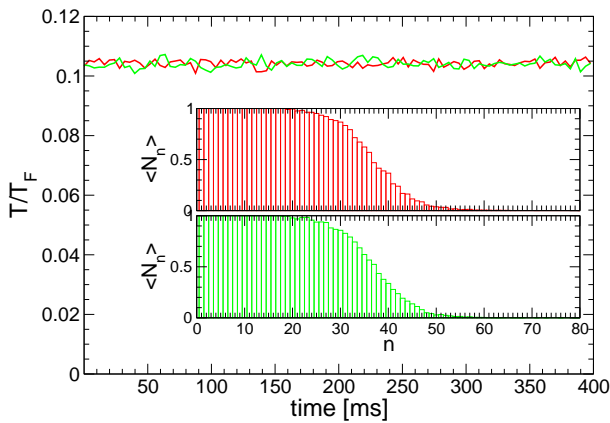}
\caption{
\label{fig:6}
Time dependence of the temperatures of the two components during the evolution in the presence of 
collisions only.
The inset presents the final distribution of the first component (upper plot) and of the second 
component (lower plot).}
\end{figure}

\begin{figure}[tb]
\includegraphics[width=8cm,clip]{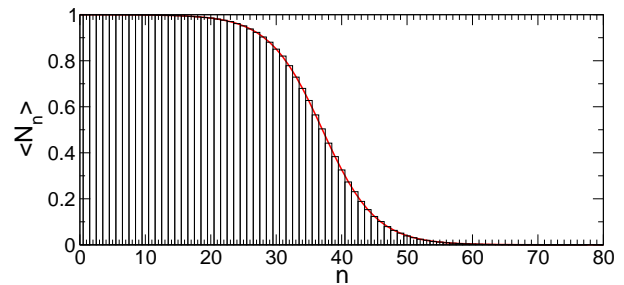}
\caption{
\label{fig:7}
Average distribution of atoms of the first component calculated during the evolution in the presence
of collisions. Solid line - fit of the thermal distribution.}
\end{figure}

\begin{figure}[tb]
\includegraphics[width=8cm,clip]{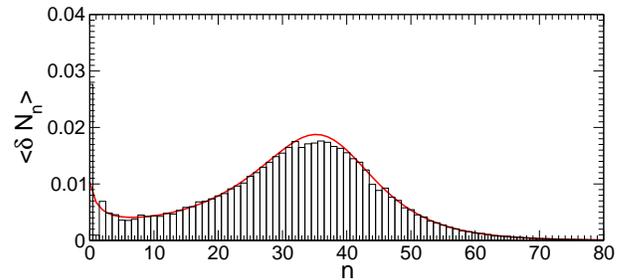}
\caption{
\label{fig:8}
Fluctuations of the atomic distribution for the first component, calculated during the evolution 
in the presence of collisions. Solid line - fluctuations of the thermal gas with the temperature and 
the chemical potential determined from the fit of the mean distribution.}
\end{figure}

Our first aim was to investigate the dynamics and the properties of the equilibrium distribution 
of the two--component Fermi gas where the collisions are described by the QBME, and the resulting
probabilities (\ref{Pcoll}). In this case we do not take into account the laser cooling 
mechanism. The energy levels in the trap are initially populated according to a thermal distribution 
with temperature $T=0.1 T_F$. Then, we let the system evolve in the presence of collisions during a time 
$t=400$~ms, corresponding to about $30$ collisions per atom in average. 
The time dependence of the temperatures of the two components is presented in Figure~\ref{fig:6},
whereas the inset shows their final distributions. As one can notice, the temperature in the system slightly 
fluctuates, but its mean value remains constant during the evolution. Therefore the final equilibrium 
distribution can not be very different from the thermal distribution. The detailed comparison may be
carried out on the basis of the mean populations and the fluctuations of the energy shells, depicted 
in Figures~\ref{fig:7} and \ref{fig:8} respectively. The bars present the averages calculated in the 
simulation, during the whole time of evolution. The solid lines depict the corresponding values
for the thermal distribution. We see, that both the mean values and fluctuations in the two--component 
system are very close to the functions calculated for a thermal distribution. This means that the QBME
correctly describes the process of thermalization. In addition 
we conclude that for $N=10660$ atoms, finite size effects are not observed, and the distributions
correspond to the ones derived in the thermodynamic limit.

\begin{figure}[b]
\includegraphics[width=8cm,clip]{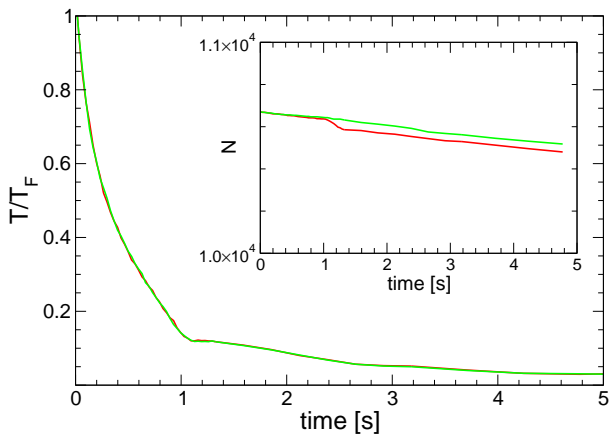}
\caption{
\label{fig:9}
Time-dependence of the temperatures of the two components (darker and lighter curves) during the cooling
of a two-component Fermi gas by controlling the effective spontaneous emission rate. 
Inset: time dependence of the number of atoms. The losses are due to background collisions and 
the removal of long-living excited atoms.
The different Raman cooling stages discussed in the text extend from $0$~$s$ to $1.3$~$s$, 
from $1.3$~$s$ to $3.2$~$s$, and from $3.2$~$s$ to $5$~$s$.}
\end{figure}

Let us now discuss the results calculated with the inclusion of the cooling mechanism. The time 
dependence of the temperature for the cooling of the two-component gas of fermions is presented in
Figure~\ref{fig:9}. The inset shows the number of fermions as a function of time. 
The initial state of the system is given by a thermal 
distribution with temperature $T_0=T_F$. As for the case of single component Fermi gas, the cooling 
process was divided into three stages, each consisting of a sequence of two Raman pulses.  
The pulses that are used are characterized by the following parameters: detuning   
$\delta/\omega=\{(-11,-12)$, $(-16,-17)$, $(-19,-20)\}$ respectively, Rabi frequency
$\Omega/\gamma=\{(0.113,0.113)$, $(0.008,0.012)$, $(0.0025,0.004)\}$ respectively, and length
$\Delta t/\omega^{-1}=\{(250,250)$, $(2000,2000)$, $(4000,4000)\}$ respectively. For the considered 
parameters, not more than $10\%$ of the atoms is excited during each pulse. As one can observe,
a final temperature $T\simeq 0.03 T_F$ may be reached within $6$~s.

\begin{figure}[t]
\includegraphics[width=8cm,clip]{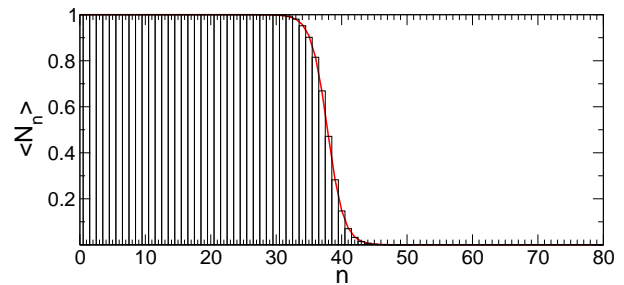}
\caption{
\label{fig:10} 
Mean final distribution of fermions (bars) obtained by averaging in time during the last stage of 
the cooling process presented in Fig.~\ref{fig:9}.
Solid line - fit of the thermal distribution.  
}
\end{figure}

\begin{figure}[t]
\includegraphics[width=8cm,clip]{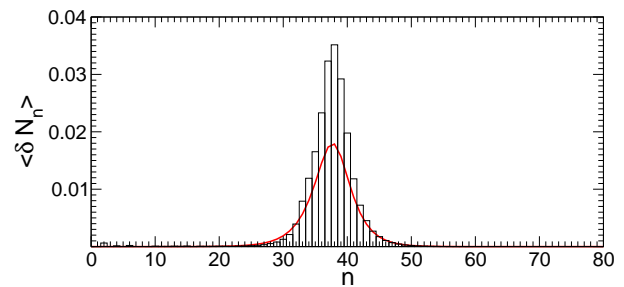}
\caption{
\label{fig:11}
Fluctuations of the final distribution of fermions (bars) 
obtained by averaging in time during the last stage of the cooling 
process presented in Fig.~\ref{fig:9}. 
Solid line - fluctuations of the thermal gas, with the temperature and the chemical 
potential determined from the fit of the mean distribution. 
}
\end{figure}

We have investigated also the statistical properties of the equilibrium distribution in the two-component 
system, in the presence of both laser-cooling and collisions. The results are presented in 
Figure~\ref{fig:10} (mean occupation numbers) and in Figure~\ref{fig:11} (fluctuations). The bars represent 
the numerical results, calculated by time--averaging during the last stage of the cooling. The solid
curves depict the corresponding functions for the thermal distribution, with the temperature and the 
chemical potential determined from the fit of the average distribution.   
The mean occupation numbers are exactly the same as predicted by the thermal distribution. The fluctuations
in the laser--cooled system, however, are substantially larger than the fluctuations in thermal ensemble.
The laser-cooling process induces additional migration of the atoms in the region close to the Fermi energy,
leading to the increased fluctuations in this region.

\begin{figure}[t]
\includegraphics[width=8cm,clip]{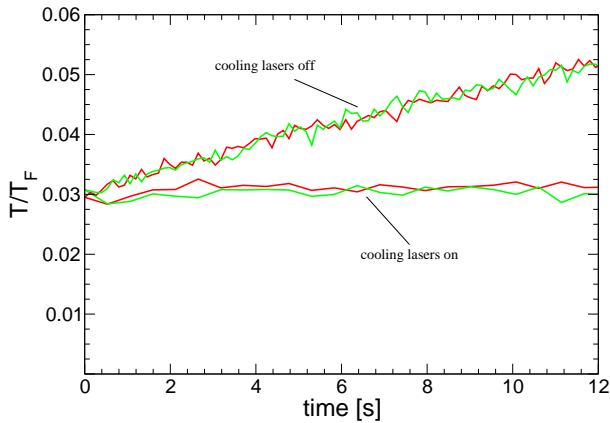}
\caption{
\label{fig:12}
Heating due to background collisions in the degenerate two--component Fermi system. 
The background collisions rate is assumed to be $\gamma=350$~Hz.
The temperature of the system slowly increases when the cooling lasers are turned off.
In the presence of the cooling lasers the temperature of the system remain unchanged.}
\end{figure}

Finally we have studied the storage of the degenerate Fermi gas produced in the cooling process.
The main limitation in this case is provided by the background losses, which may generate 
holes deep within the Fermi sea. The latter increases the corresponding energy per particle, and 
via collisional thermalization leads to an important source of heating \cite{Timmer}. 
We calculate the dynamics of 
the system, which was already cooled to a temperature $T=0.03 T_F$. The time dependence of the 
temperature is presented in Figure~\ref{fig:12}. The plot compares two cases: (i) the 
laser is turned off, and the gas is heated due to the background collisions, (ii) the laser is 
turned on, and the cooling pulses of the last stage are continuously applied. As we may observe from the
figure, in the latter case, the laser cooling compensate for the heating induced by the creation of 
holes in the degenerate distribution. Hence, it helps to maintain the degenerate gas for a relatively 
long time in the trap. We have also verified that the presence of laser-cooling does not lead 
to substantially larger losses.

\section{Conclusion}

In this paper we have analyzed the laser cooling of trapped one and two-component Fermi gases.
We have developed the quantum 
master equation that describes the laser cooling of fermions, and
the process of collisions in the case of the two-component gas. 
We have considered the FL regime where the heating due to photon reabsorptions can be neglected.
We have shown that the inhibition of the spontaneous emission due to Pauli blocking  
can be overcome employing collective Raman cooling with dynamically modified spontaneous Raman rate.
Additionally, a slightly anharmonic trap could be employed. By means of Monte Carlo 
simulations of the corresponding rate equations, we have shown that 
it is possible to cool one- and two-component Fermi gases down to temperatures of the 
order of a few percent of $T_F$. The results obtained in this paper present by no means fundamental 
limits of laser cooling; temperatures below 1\% of $T_F$ can be achieved 
by optimizing the cooling procedure \cite{TwoComp}.

The laser cooling is an external process that influences the dynamics of trapped atoms, therefore one can
expect that the equilibrium distribution of the laser-cooled system may differ from the thermal distribution.
We have analyzed, both for the single- and the two-component case, the statistics of the laser-cooled
gas, comparing the results to the predictions of the grand canonical ensemble. 
We have shown that although the mean-number distribution can be well approximated by a thermal one, 
the fluctuations are significantly different. In particular, the fluctuations are considerably enhanced 
at the Fermi surface. This effect could play a significant role in the BCS pairing in laser cooled gases, 
and it will be the subject of further analysis.

Additionally, we have discussed the possible loss sour\-ces that may affect the laser cooled gas.
In particular, for the two-component gas, the creation of holes deep in the Fermi sea due to background 
collisions could lead to a significant heating of the sample, which should importantly reduce the life-time 
of the degenerate gas. In this sense, we have shown that our laser-cooling scheme 
can be employed to maintain a two-component  
Fermi gas at a fixed temperature without significant additional atom losses.

\begin{acknowledgments}

We acknowledge support from the Alexander von Humboldt Stiftung, 
the Deutsche Forschungsgemeinschaft, the RTN ``Cold Quantum Gases'', 
the ESF Program BEC2000+, the Russian Foundation for Fundamental Research, 
the Polish KBN Grant No. 5-P03B-103-20, and IST Program ``EQUIP''.  

\end{acknowledgments}

\end{document}